\documentstyle[aps,epsfig,multicol]{revtex}

\def\up{\uparrow}
\def\dn{\downarrow}

\def\rt{\rightarrow}
\newcommand{\bup}{\Uparrow}
\newcommand{\bdn}{\Downarrow}

\def\brt{\longrightarrow}

\def\beq{\begin{equation}}
\def\eeq{\end{equation}}

\begin{document}

\title{Arrested States formed on Quenching Spin Chains with Competing Interactions and Conserved Dynamics}
\author{Dibyendu Das and Mustansir Barma}
\address{Department of Theoretical Physics, Tata Institute of Fundamental Research,
Homi Bhabha Road, Mumbai 400 005}
\maketitle

\begin{abstract}

We study the effects of rapidly cooling to $T = 0$ a spin chain with
conserved dynamics and competing interactions. Depending on the degree
of competition, the system is found to get arrested in different kinds
of metastable states. The most interesting of these has an
inhomogeneous mixture of interspersed active and quiescent regions. In
this state, the steady-state autocorrelation function decays as a
stretched exponential $\sim exp(-{(t/\tau_{o})}^{{1}\over{3}})$,
and there is a two-step relaxation to equilibrium when the 
temperature is raised slightly.


\end{abstract}
\begin{multicols}{2}

When a system at a high temperature is cooled rapidly to low
temperatures, it may not be able to reach an equilibrium state in an
experimentally realizable time. Instead it may reach a long-lived arrested 
state, often with a degree of spatial disorder, which ultimately relaxes
towards equilibrium over very much longer time scales.
An intriguing possibility is that the disorder
induced by the kinetics may be strong enough that the system has widely
different levels of dynamical activity in distinct regions of space.
Dynamically heterogeneous states are found to arise, for instance, 
in a glass-forming liquid \cite{donati}. From the theoretical point of
view it is important to ask: Are there simple models in which dynamically
heterogeneous states arise naturally? Can their formation and properties  
be understood in microscopic terms? Finally how do such
states decay, and how is equilibrium approached?

We address these questions by studying nonequilibrium quenches to $T = 0$,
in simple lattice models. Following such 
quenches, the system may get arrested in a metastable state instead of 
reaching the ground state.      
A useful way to characterize the resulting arrested state is to ask
whether or not there is any dynamical activity in it. A {\it quiescent} 
arrested state is one in which the system settles into a single configuration, 
and degrees of freedom are frozen. Another possibility is that the arrested 
state may involve a large number of 
configurations which are dynamically accessible from each other; in that 
case the system is dynamically {\it active}. Interestingly, all these 
possibilities are realized in quenches of a family of simple models, namely
Ising chains with different degrees of competition and conservation laws.
In the absence of competition, the system approaches the ground state if the  
dynamics is nonconserving \cite{Bray}, while it reaches a quiescent arrested 
state under 
spin-conserving dynamics \cite{Priv}. On the other hand, a system 
with competing interactions, evolving under nonconserved dynamics, has been 
shown to exhibit an active arrested state \cite{redner}. This naturally 
leads to the question: Are new features brought in if both conservation and 
competition are present? 
In this Letter, we study the effects of quenching 
the simplest model which incorporates both these features,
namely an Ising model with competing first and second neighbour interactions,  
evolving through a 
dynamics with a single conservation law. Despite its simplicity, the model 
shows interesting transitions in the character of the arrested states  
as the degree of competition is varied (Fig. 1).
The most interesting of these arrested states, reached for strong enough 
competition, is of a qualitatively new type: it has active and quiescent 
regions interspersed in a disordered fashion throughout the system. This
{\it inhomogeneous quiescent and active} (IQA) state, has nontrivial
dynamical properties. In the $T = 0$ steady state the
autocorrelation function has a stretched
exponential decay. Further if the temperature is raised slightly, there
is a two-step relaxation: the IQA state relaxes to equilibrium via an 
intermediate long-lived intermediate energy state. We are able to
quantitatively understand many of these unusual features, often found in 
glassy systems, within this simple model.

The equilibrium phases and transitions of the Axial Next Nearest 
Neighbour Ising (ANNNI) model have been well studied and characterized 
\cite{yeomans}. However, its nonequilibrium properties 
remain relatively unexplored even in 1-d, except for a few studies. The
first such study explored arrested states obtained by quenching across $T = 0$
phase boundaries of an extended ANNNI model appropriate to polytypes
\cite{kabra}. More recently, time-dependent coarsening induced by
nonconserved dynamics has also been studied by quenching the system
across phase boundaries at $T=0$ \cite{shreshtha}, and also from
$T=\infty$ to $T=0$ \cite{redner}. Here we explore the interplay of 
competing interactions with conservation laws in the dynamics.
Our principal result is the identification and characterization
of an IQA arrested state in this system.

Consider an Ising chain with spin variables $\{{s_{i}}\}$ described by
the Hamiltonian
\beq
{\mathcal{H}}=-{J_{1}}{\sum}_{i}{s_{i}}{s_{i+1}} + {J_{2}}{\sum}_{i}{s_{i}}{s_{i+2}} .
\label{eq:hamil}
\eeq
An antiferromagnetic next neighbour coupling $({J_{2}} > 0)$ competes
with the nearest neighbour coupling ${J_{1}}$ which may be of either
sign. In what follows, we shall assume ${J_{1}} > 0$ (ferromagnetic
coupling) and define ${j_{2}} = {J_{2}}/{J_{1}}$ as a measure of the
strength of competition. The equilibrium ground state shows a
transition from the ferromagnetic state $\up\up\up\up...$ for ${j_{2}}
< 0.5$, to the antiphase state $\up\up\dn\dn...$ for ${j_{2}} >
0.5$. The point ${j_{2}} = 0.5$ is a multiphase point, at which the
number of ground states (all configurations with no single spins)  
is exponentially large in system size $L$
\cite{yeomans}.  We are interested in the effect of a quench from an
infinite-temperature random configuration of the system to $T = 0$.
We use a double-spin-flip dynamics (DSFD), in which an adjacent 
pair of randomly chosen parallel spins is flipped 
($...\up\up... \brt ...\dn\dn...$).  Flips are attempted at unit rate, 
and are allowed only if the energy is not raised $({\Delta E}{\leq} 0)$.
Evidently, the DSFD conserves the difference $M = M_{1} - M_{2}$ of
the two sublattice magnetizations $M_{1}$ and $M_{2}$. 
The dynamics thus involves a single conservation law. 
The DSFD maps onto the well-known Kawasaki spin exchange
dynamics through a sublattice mapping, in which every spin on one of
the two sublattices is inverted and the sign of the nearest neighbour
coupling ${J_{1}}$ is reversed. In the mapped model with Kawasaki dynamics, 
$M$ is the total conserved magnetization. We will use the DSFD,
rather than Kawasaki description; results can be translated readily.
The DSFD can be looked upon as an extension of
the single-spin-flip Glauber dynamics to multiple-spin flips at a
time. Multispin moves arise in physical contexts such as
stacking dynamics in the 3C-6H transition in $SiC$ \cite{kabra2}
and deposition-evaporation dynamics of bunches of particles \cite{stinch}.

The energy non-raising condition, which is a consequence of a $T = 0$ 
quench, imposes local constraints on whether
or not a pair of chosen spins can actually be flipped; these
constraints are a function of ${j_{2}}$. The normalized energy changes
${\Delta e}{\equiv} {\Delta E}/{J_{1}}$ involved in flipping a pair
$\up\up$ to $\dn\dn$ depend on the environments of the pair, and are
given in Eq.\ref{eq:en-moves} below. There are six distinct local
environments; the other unlisted environments are related to these by
reflection symmetries.
\begin{eqnarray}
	(a) ~~{\bup}{\bdn}{\up}{\up}{\bdn}{\bdn} &{~\brt~}& {\bup}{\bdn}{\dn}{\dn}{\bdn}{\bdn} \hskip 0.5cm {\Delta e}=-(4-4{j_{2}}) \nonumber \\
	(b) ~~{\bdn}{\bup}{\up}{\up}{\bup}{\bdn} &{~\brt~}& {\bdn}{\bup}{\dn}{\dn}{\bup}{\bdn} \hskip 0.5cm {\Delta e}=4 \nonumber \\
	(c) ~~{\bdn}{\bup}{\up}{\up}{\bdn}{\bdn} &{~\brt~}& {\bdn}{\bup}{\dn}{\dn}{\bdn}{\bdn} \hskip 0.5cm {\Delta e}=4{j_{2}} \nonumber \\
\label{eq:en-moves}
	(d) ~~{\bdn}{\bdn}{\up}{\up}{\bdn}{\bdn} &{~\brt~}& {\bdn}{\bdn}{\dn}{\dn}{\bdn}{\bdn} \hskip 0.5cm {\Delta e}=-(4-8{j_{2}}) \\
	(e) ~~{\bdn}{\bup}{\up}{\up}{\bdn}{\bup} &{~\brt~}& {\bdn}{\bup}{\dn}{\dn}{\bdn}{\bup} \hskip 0.5cm {\Delta e}=0 \nonumber \\
	(f) ~~{\bdn}{\bdn}{\up}{\up}{\bup}{\bup} &{~\brt~}& {\bdn}{\bdn}{\dn}{\dn}{\bup}{\bup} \hskip 0.5cm{\Delta e}=0 \nonumber 
\end{eqnarray}
The reverse of move $(a)$ in $(\ref{eq:en-moves})$ will be
referred to as $(\bar{a})$, and similarly for the others.

Evidently the dynamics is identical for all values of
${j_{2}}$ for which the same set of moves are allowed. Moves
$(\bar{b})$, $(e)$, $(\bar{e})$, $(f)$ and $(\bar{f})$ are allowed
(i.e. ${\Delta e} \leq 0$) for all ${j_{2}}$. As ${j_{2}}$ is varied, 
${\Delta e}$ changes sign for moves $(a)$, $(c)$ and $(d)$. Each such change
causes a change in the nature of the arrested state. While $(c)$ is
an allowed move for ${j_{2}} < 0$, $(\bar{c})$ becomes an allowed move for
${j_{2}} > 0$. Similarly, across ${j_{2}} = 0.5$ and ${j_{2}} = 1$, the 
allowed moves change from $(d)$ to $(\bar{d})$ and $(a)$ to $(\bar{a})$,
respectively.  Thus there are distinct regions of dynamical activity 
along the ${j_{2}}$ axis: (i) ${j_{2}}{\in}(-{\infty},0)$,
(ii) ${j_{2}}{\in}[0,0.5)$, (iii) ${j_{2}} = 0.5$, (iv) ${j_{2}}{\in}(0.5,1]$,
and (v) ${j_{2}}{\in}(1,{\infty})$; see Figure 1. In region (v), i.e. 
for strong competition, the system reaches an IQA 
arrested state.

We used Monte Carlo simulation to study the arrested steady states
that are reached under ${\Delta e}\leq 0$ DSFD starting from a random
initial configuration corresponding to $T=\infty$. We studied the 
approach to the arrested states, the dynamical behavior in these states, and 
finally the relaxation from these states to equilibrium at low but finite 
temperatures. The approach to, and the decay from, the steady state was 
monitored by following the decay of the energy in time. Further, 
in cases (iii) and (v), we studied the dynamical behavior of the steady 
state by monitoring the spin-spin autocorrelation function
\beq
C(t) = {1 \over N}{\sum}_{i} {\langle}s_{i}(t_{o}) s_{i}(t_{o} + t){\rangle} - {\langle}s_{i}(t_{o}){\rangle}^{2}
\label{eq:corr}
\eeq
where $t$ is the number of Monte Carlo steps per spin and
$\langle ... \rangle$ denotes an average over ${t_{o}}$. We 
allowed for an explicit dependence of averages on the space 
location $i$, as arrested states need not be translationally
invariant. 
Only at the multiphase point (iii) is the ground 
state reached on quenching; in the other four regions of ${j_{2}}$ discussed 
above (Fig. 1), the steady states are arrested.

Before discussing the IQA state in detail, we 
sketch some features of the states in the other four regions. 

\noindent~(i)~The arrested state is quiescent. It consists of 
ferromagnetic patches separated by clusters of frozen domain walls, 
e.g. $...\up\up\up\up\dn\up\dn\up\up\up...$. It is qualitatively similar
to the arrested state obtained in \cite{Priv}, with only first neighbour
interactions.

\noindent~(ii)~The steady state has a number 
of diffusing domain walls separating ferromagnetically aligned patches 
(Fig. 1). Though it resembles the active arrested states found in 
\cite{redner}, there is an important difference. The level of activity is 
much lower in our case, as the number of walls increases as 
$\sim L^{1 \over 2}$ as opposed to $\sim L$ in \cite{redner}.

\noindent~(iii)~At the multiphase point there is a large degree of 
activity (Fig. 1), because the ${\Delta e} = 0$ moves (d), (f) and their 
reverses carry the system through a subspace of ground state 
configurations labeled by a given value of $M$. The autocorrelation 
function $C(t)$ $\sim t^{-{1 \over 2}}$ at long times, as in the  
unconstrained DFSD \cite{stinch}.

\noindent~(iv)~The steady state has alternating opposite-spin clusters 
of $2$ or $3$ spins, e.g. $...\dn\dn\dn\up\up\dn\dn\dn\up\up\up...$, 
and is quiescent. A single cluster of 
$4$ or $5$ spins may remain in the steady state and diffuse through 
a quiescent background (Fig. 1).

In region (v), an IQA arrested state with alternating quiescent and 
active stretches, is reached. A segment of 
a typical configuration is depicted below.  
\begin{eqnarray*}
..\left[\bdn\bup\bup\bdn\bdn\bup\right]&\dn\up\dn\dn\dn\up\dn\up\dn\up\up\up\dn\up\dn&\left[\bup\bdn\bdn\bup\bup\bup\bdn\bdn\bup\bdn\bdn\bup\bup\bup\bdn\bdn\bup\right]..\\
 {\rm Quiescent}     & {\rm Active}      &  {\rm Quiescent} \\  
\end{eqnarray*}
Each active region has parallel-spin triplets in a background of alternating 
single spins, while the quiescent portions predominantly resemble the arrested 
state of region 
(iv). Crucial to the coexistence of active and quiescent regions is the 
existence of stable walls at the boundaries of quiescent regions. These 
consist of left boundaries $\bup\bdn\bdn$ or $\bdn\bup\bup$ and right 
boundaries $\bdn\bdn\bup$ or $\bup\bup\bdn$, and are stable as  
moves $(a)$ and $(c)$ are energy-raising for $j_{2} > 1$.
The numbers of quiescent and active ($q$ and $a$ respectively) regions
of size $\tilde\ell$
are found numerically to decay as $exp(-\lambda \tilde\ell)$ with 
$\lambda_q \simeq 0.05$ and $\lambda_a \simeq 0.25$.

We now turn to the dynamical properties of the IQA state. The {\it 
autocorrelation 
function} in the steady state decays as a stretched exponential $\sim
exp(-{(t/\tau_{o})}^{1 \over 3})$ (Fig. 2).  
Interestingly, the dynamical behavior of the IQA state can be related to
the well-known symmetric exclusion process (SEP) of particles on a line 
\cite{liggett}. This 
can be understood as follows.  Each spin triplet in an active stretch 
can move by one unit right or left, under the DSFD move $(e)$ 
($...\dn\up\dn{\overline{\up\up\up}}\dn\up...\longleftrightarrow...\dn\up{\underline{\dn\dn\dn}}\up\dn\up...$). 
There is a hard-core repulsion between triplets as move $(c)$ is 
disallowed. The dynamics within a single active region is thus precisely 
that of a symmetric exclusion process of hard triplets on a lattice, where
the single spins can be viewed as holes. 
The autocorrelation function $C_{\tilde\ell}(\rho,t)$ 
averaged over spins, of an active stretch of length $\tilde\ell$ 
with $\rho\ell$ triplets (where $\ell = \tilde\ell + 2$ and $1/\ell \leq \rho 
\leq 1/3$), is thus governed by the diffusion of these hard triplets;  
triplets extend over an extra lattice unit at both boundaries of an active
stretch making its effective length $\ell$. Hence, we expect
$C_{\tilde\ell}(\rho,t)$ to decay as $t^{-{1 \over 2}}$ for times $t$ less
than a cutoff time $\tau_{\ell}(\rho)$, and as
$exp(-t/\tau_{\ell}(\rho))$ thereafter.
Further,  $\tau_{\ell}(\rho)$ can be found by noting an exact mapping of 
every configuration of this
problem to a corresponding configuration of the SEP. Under the mapping,
every triplet is replaced by a single particle, while a
single spin maps onto a hole ($\up\dn\up\dn\dn\dn\up\dn\up\up\up\dn\dn\dn\up\dn
~\brt~\circ\circ\circ\bullet\circ\circ\bullet\bullet\circ\circ$). 
The mapped chain has a reduced length $\ell^{\prime} = {\ell}(1 - 2\rho)$. 
The stochastic
${\mathcal{W}}$-matrices for the two processes are same, as there is a
1-1 correspondence between configurations and moves.  This
implies that the eigenvalue spectra of the ${\mathcal{W}}$-matrices in the 
two problems are the same, and in particular, the gap $\Delta_{\ell}$ to the 
first excited state is the same. The inverse of the gap is just the cutoff time
$\tau_{\ell}$, and so the above equality implies $\tau_{\ell} =
\tau_{\ell^\prime}^{\prime}$. For the exclusion process, with free
boundary conditions $\tau_{\ell^\prime}^{\prime} = 2
{\ell^\prime}^{2}/\pi^{2}$ for large ${\ell^\prime}$ (the diffusion constant
$= {1 \over 2}$) \cite{tau_l}, and 
hence $\tau_{\ell} (\rho) = 2 {\ell}^{2}(1 - 2\rho)^{2}/\pi^{2}$.

The autocorrelation function $C_{IQA}(t)$ of the IQA state can be 
expressed in terms of a sum over active stretches:
\beq
C_{IQA}(t)=\sum_{\tilde\ell,\rho} P_{\tilde\ell}(\rho) C_{\tilde\ell}(\rho,t)
\label{eq:colcorr}
\eeq  
where $P_{\tilde\ell}(\rho)$ is the probability of finding an active stretch
of length $\tilde\ell$ and density $\rho$ of triplets. 
Even without explicitly determining  $P_{\tilde\ell}(\rho)$, we can derive 
bounds 
on  $C_{IQA}(t)$ using $P_{\tilde\ell} = \Sigma_{\rho} P_{\tilde\ell}(\rho) 
\sim exp(-\lambda_{a} \tilde\ell)$.
As $\rho$ varies across its range, the cutoff time $\tau_{\ell}$ 
varies between the two limits $\tau_{\ell} (0) = 2 {\ell}^{2}/\pi^{2}$ (for a
single triplet) and $\tau_{\ell}({1\over 3}) = 2 {\ell}^{2}/ 9 {\pi}^{2}$ 
(for a single hole hopping over $3$ lattice units). For each of these limits
$\tau_{\ell}(\rho^{*})$, the sum in Eq.(\ref{eq:colcorr}) is
dominated at long times by the term with the saddle point value 
$\ell^* = {({{t \pi^{2}}/{\lambda_{a} (1 - 2\rho^{*})^{2}}})}^{1 \over 3}$. 
The bounds imply that $C_{IQA}$ has a stretched exponential form 
$\sim exp(-{(t/\tau_{o})}^{1 \over 3})$, with 
${8/{243\pi^{2}}} \leq \tau_{o}\lambda_a^{2} \leq {8/{27\pi^{2}}}$.
The numerically determined values $\tau_{o} \simeq 0.08$ and
$\lambda_a \simeq 0.25$ are consistent with these bounds 
(Fig. 2).

The {\it dynamics of approach} to the IQA state, starting from a random 
initial 
configuration, is also interesting. From numerical simulations the energy
is found to decay as $\sim exp(-(t/t_{o})^{1 \over 3})$ (see inset in 
Fig. 2).   
This is associated primarily with the fall in the number $N_4$ of $4$-spin 
clusters which diffuse through ground state stretches $...\up\up\dn\dn...$
until they dissociate when they encounter `traps' in the form of
a single spin or a triplet; e.g. $...\up\up\dn\up\up\up\up... \rt 
...\up\up\dn\dn\dn\up\up...$.
The typical time for a $4$-cluster to diffuse over 
a length $l$ before encountering 
such a trap is $l^{2}/D$, implying that $N_4(l)$ decays as 
$exp(-Dt/l^{2})$. Further, the stretch lengths 
are distributed exponentially ($\sim exp(-\lambda l)$), so that the 
average of $N_4(l)$ over $l$ is dominated by a saddle point value 
$l^* = (2Dt/\lambda)^{1 \over 3}$ at  
large $t$. This argument is reminiscent of that in \cite{cornell} and implies 
a stretched exponential form for the decay.

Finally, let us discuss the {\it relaxation to equilibrium} from the IQA 
arrested state. The IQA state has regions 
with two distinct types of excitations, namely active patches with mostly 
single spins (and occasional mobile triplets) and quiescent patches with 
mostly triplets (and occasional frozen single spins). If $T$ is
raised to a small value, the system relaxes to 
an equilibrium state close to the $...\up\up\dn\dn...$ ground 
state, by annealing out  both the
single-spin and triplet excitations. Figure 3
shows the subsequent variation of energy with time. There is
a relatively rapid approach to a second metastable state, evidenced 
by a long plateau (Fig. 3), followed by an eventual approach to
equilibrium. This can be understood as follows. For finite $T$,
the energy raising moves $(a)$, $(b)$, $(c)$ and $(d)$ are  
allowed, with probabilities $\omega_{k} \sim exp(-{\Delta e}_{k}/T)$,
$k$ = $a$,$b$,$c$,$d$; the associated time scales are $\sim
1/\omega_{k}$. Moves $(a)$ and $(c)$ are instrumental in annealing 
out the two different kinds of local excitations (isolated spins and 
triplets respectively). They 
act on time scales which are widely different ($\omega_{c}^{-1}/\omega_{a}^{-1}
\sim exp(4{J}_{1}/T)$) leading to the plateau. 
For  small $t$ , i.e. $t \sim \omega_{a}^{-1}$, 
only move (a) is effective, which destabilizes active-quiescent 
boundaries and creates $5$-spin clusters. 
Single spins diffuse out of active stretches
and annihilate on meeting $5$-clusters e.g. $\dn\dn\up\dn\dn\dn\dn\dn
 \rt \dn\dn\up\up\up\dn\dn\dn$. After the single spins anneal out, the
system reaches a metastable state with clusters of length $2$ and $3$, much
like the arrested state in region (iv).  
This continues till $t \stackrel{>}{_\sim} {\omega}_{c}^{-1}$, 
when triplets begin to decay.
To leading order in low $T$, the predominant decay channel involves
the following steps: $(i)$ the conversion of a triplet to a single spin and 
a $4$-spin cluster (at rate $\omega_c$), $(ii)$ the production of a 
single spin when the $4$-cluster meets the nearest triplet, and $(iii)$ the 
fast diffusion of single spins till they meet triplets or single spins at 
separation $2n$ ($n$ = odd), whereupon they annihilate e.g.
$\dn\dn\up\dn\dn\up\up\up \rt \dn\dn\up\up\up\up\up\up \rt 
\dn\dn\up\up\dn\dn\up\up$. Process 
$(iii)$ is a variant of the single species 
diffusion-annihilation process \cite{a+a}, 
implying a power-law ($\sim t^{-{1 \over 2}}$) decay for the energy.

To summarise, a simple understanding of the dynamics of the IQA state can be 
achieved in terms of diffusing excitations; the nature of approach, 
steady-state autocorrelation function and decay of the state involve 
variants of the diffusion problem. For instance, the approach to the IQA state
involves diffusion in the presence of randomly placed traps, while the 
autocorrelation function involves the consequences of confinement of 
diffusing excitations in active stretches of random lengths. In both cases,   
an average over the dynamically generated randomness results in a stretched 
exponential decay. The two distinct time scales for relaxation from the IQA  
state arise from the different activation rates for the two types of diffusing 
excitations.  Diffusion-limited annihilation of the second type governs the
power-law decay towards equilibrium.

We conclude by pointing out that IQA arrested states occur in several other 
situations, for instance, with antiferromagnetic nearest-neighbour
coupling (${J_{1}} < 0$), and also under a quench from a
quiescent arrested state in region (i). Further an IQA state is found in 
quenches of an extended ANNNI model relevant to polytype transitions
\cite{kabra2}. This model is richer, and shows variability in the 
microscopic nature of activity in IQA arrested states for different parameter 
values \cite{barma}. Interestingly, despite this variation, the 
dynamical behavior remains of the same form --- a general consequence of the 
diffusion-based description given above.

Acknowledgements: We thank A. Dhar, D. Dhar and S.N. Majumdar for useful
discussions.

\end{multicols} 

{\bf Figure Captions}

\noindent{Figure 1: Space-time depiction of the activity (shown white) 
in different arrested states. Different regimes: 
(i) Quiescent (ii) Mobile domain walls (iii) Active (iv) Essentially 
quiescent (v) IQA state.}

\noindent{Figure 2: The autocorrelation function in the IQA state with 
$L=12000$ and $10^{6}$ histories. The dotted curves are the 
bounds discussed in the text. Inset: Decay of energy excess (in units of
$J_1$) over the  IQA value, starting 
from a random state.}

\noindent{Figure 3: Energy per site $e(t)$ ($\bullet$) in units of $J_1$, 
measured from the ground state value, when the IQA state is taken to 
$T \simeq 0.29 {J_{1}}$. We used $\omega_{a}
= 1$, $\omega_{c} = {{0.9} \times {10^{-6}}}$, $L = 1200$ 
and $12$ initial conditions. Also shown are the 
fraction of single spins ($\triangle$) and triplets ($\Box$).}





\begin{thebibliography}{99}


\bibitem{donati}{C. Donati {\it et al.}, cond-mat$ /9810060 $}. 

\bibitem{Bray}{A. J. Bray, Adv. Phys. {\bf 43}, 357 (1994)}.

\bibitem{Priv}{R. G. Palmer and H. L. Frisch, J. Stat. Phys. {\bf 38}, 
867 (1985); V. Privman, Phys. Rev. Lett. {\bf 69}, 3686 (1992)}. 


\bibitem{redner}{S. Redner and P.L. Krapivsky, J. Phys. A: Math. Gen. 
{\bf 31} 9229 (1998)}. 


\bibitem{yeomans}{J. Yeomans, Solid State Physics {\bf 41}, 151 (1988)}. 


\bibitem{kabra}{V.K. Kabra and D. Pandey, Phys. Rev. Lett. {\bf 61}, 
 1493 (1988)}.

\bibitem{shreshtha}{S.P. Shreshtha and D. Pandey, Europhys. Lett. {\bf
34}, 269 (1996)}.

\bibitem{kabra2}{V.K. Kabra, D. Pandey and S. Lele, J. Mat. Sci. 
{\bf 21}, 1654 (1986). Double-layer displacements discussed 
there correspond to triplet flips in  H$\ddot{\rm{a}}$gg-Ising notation.}


\bibitem{stinch}{R.B. Stinchcombe, M.D. Grynberg and M. Barma, Phys. Rev. E 
{\bf 47}, 4018 (1993).}


\bibitem{liggett}{T.M. Liggett, {\it Interacting Particle Systems},
(Springer-Verlag, New York, 1985).}


\bibitem{tau_l}{ SEP correlation functions with free boundary conditions can 
be computed using the methods of \cite{stinch}.}

\bibitem{cornell}{S.J. Cornell, K. Kaski and R.B. Stinchcombe, Phys. Rev. B
{\bf 44}, 12263 (1991)}. 

\bibitem{a+a}{K. Kang and S. Redner, Phys. Rev. A {\bf 32}, 435 (1985)}.


\bibitem{barma}{D. Das and M. Barma, Physica A {\bf 270}, to appear (1999)}.
 
 
\end{thebibliography}
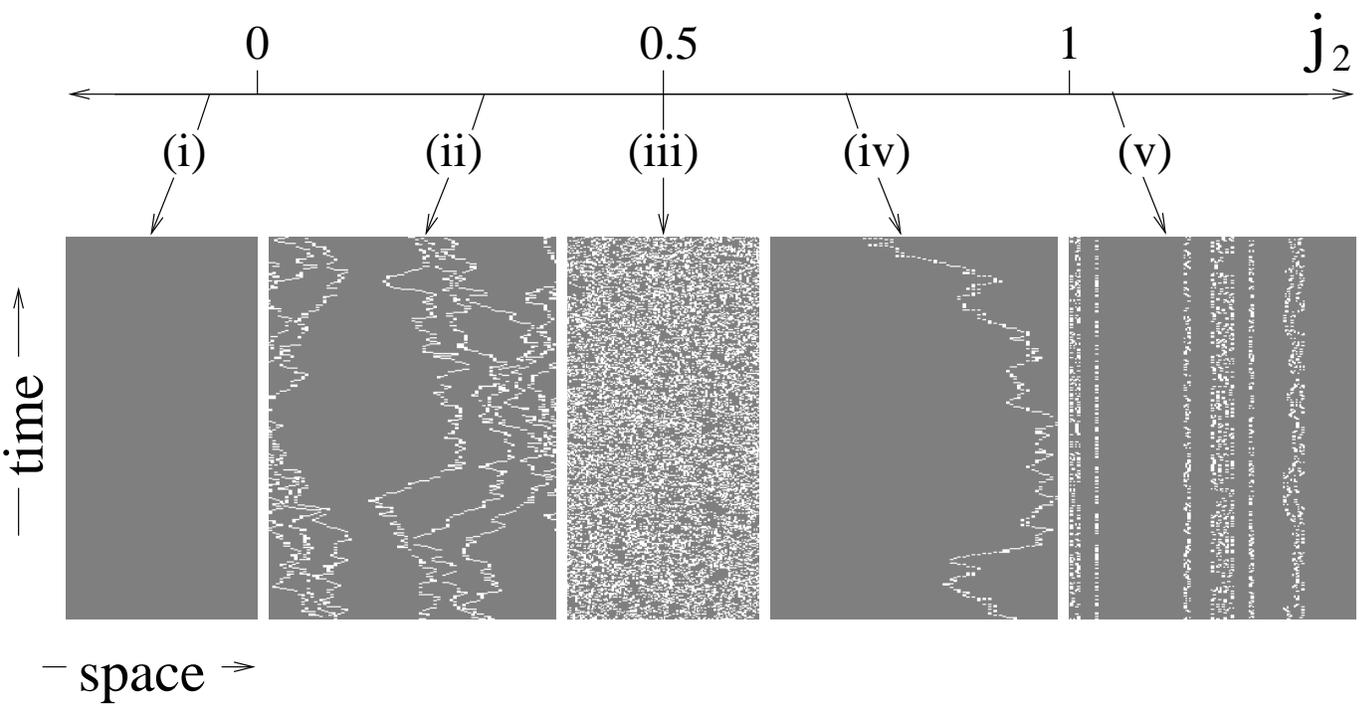
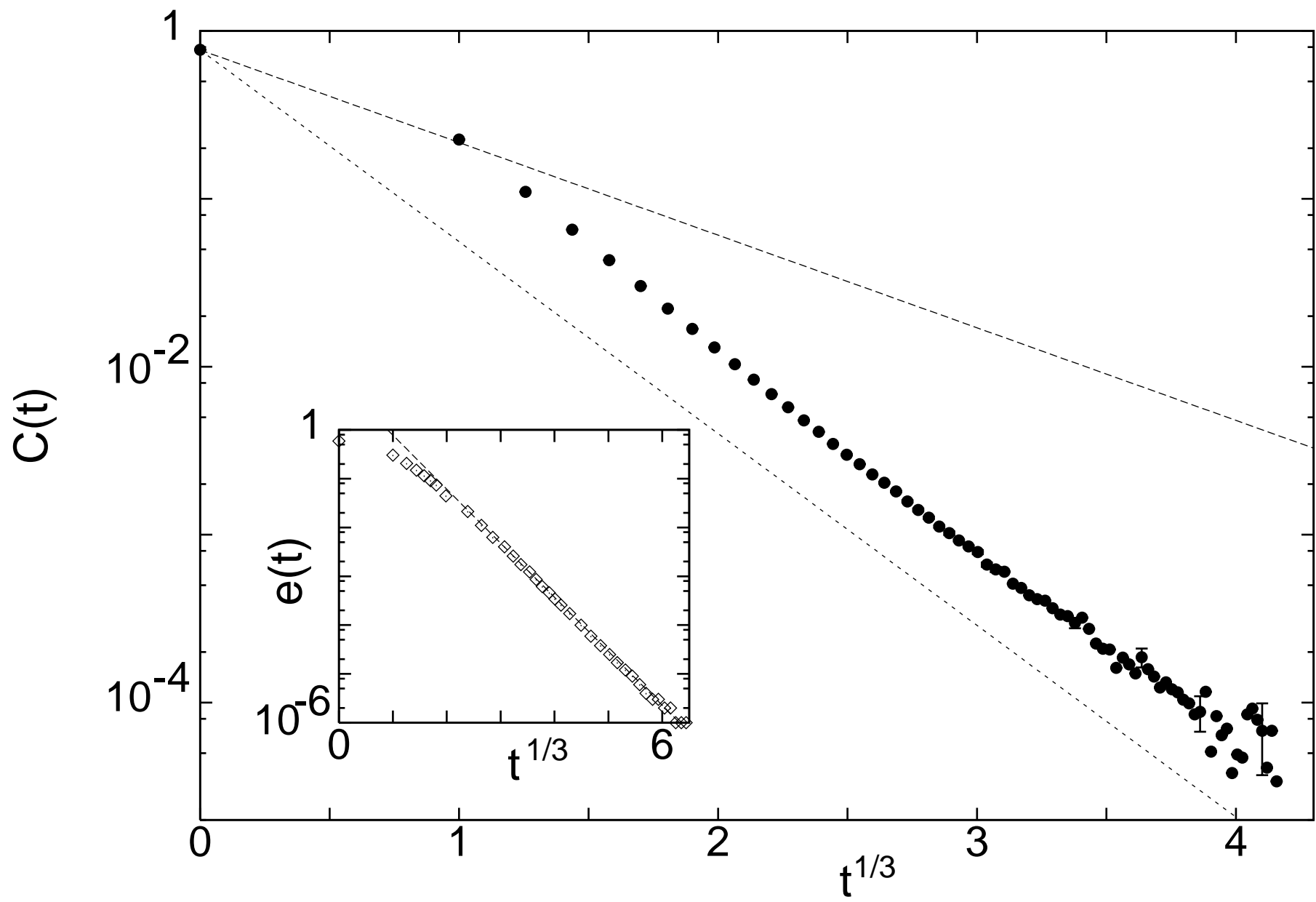
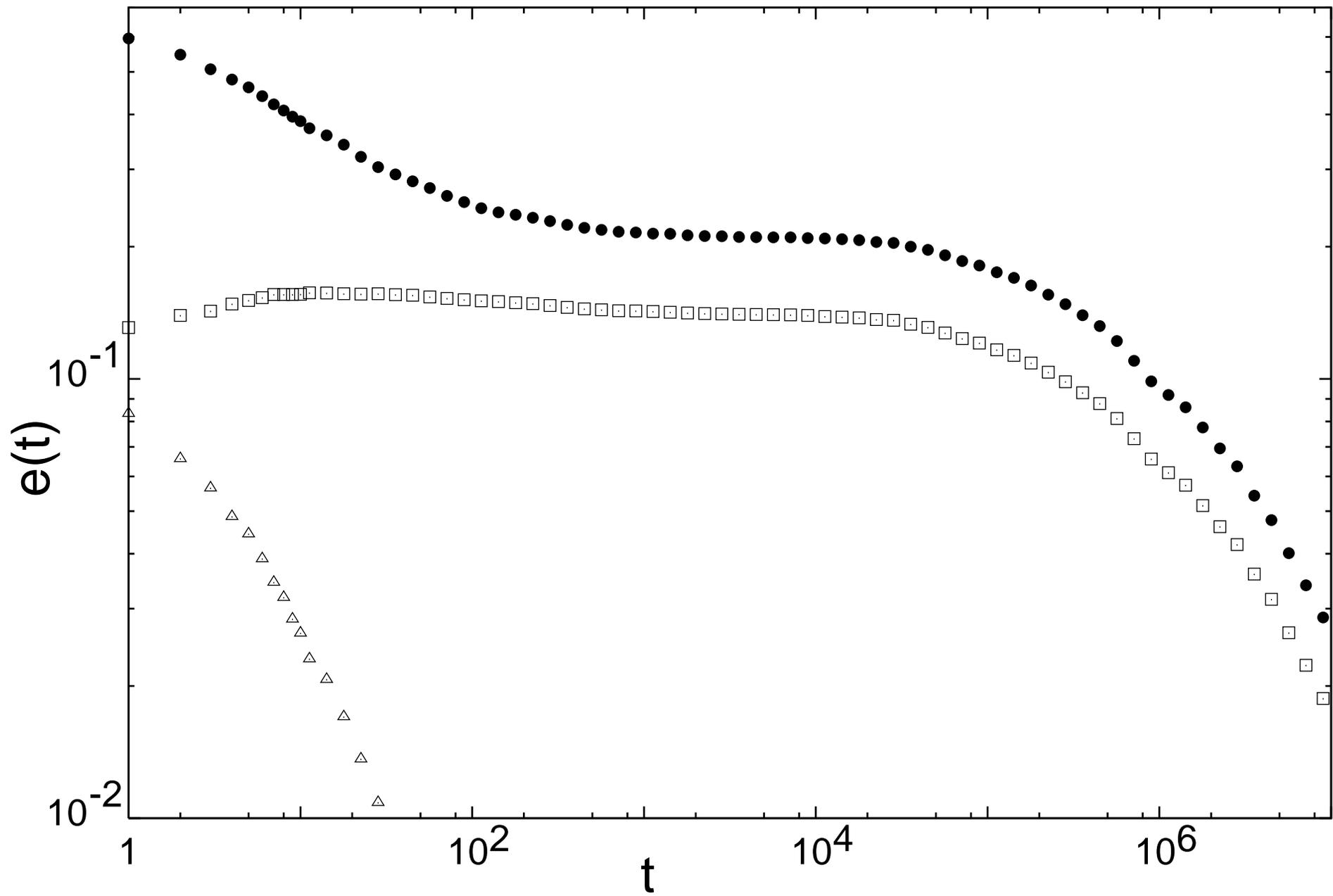
\end{document}